# Energy Transfer Controlled by Dynamical Stark Shift in Two-level Dissipative Systems

A. V. Ivanov,*
*Research Center "Information Optical Technologies",
ITMO University, Birzhevaya liniya, 14, Saint Petersburg, 199034, Russia*

A strong electromagnetic field interacting with an electron system generates both the Rabi oscillations and the Stark splitting of the electron density. Changing of the electron density gives rise to nonadiabatic effects due to existence of the electron-vibrational interaction in a dissipative system. In this Letter, the mechanism of energy transfer between the electron system and the phonon reservoir is presented. This mechanism is based on establishment of the coupling between the electron states dressed by the electromagnetic field and the forced vibrations of reservoir oscillators under the action of rapid changing of the electron density with the Rabi frequency. The photoinduced vibronic coupling results in appearance of the states that are doubly dressed by interaction, first time due to the electron-photon interaction, and second time due to the electron-vibrational interaction. Moreover, this coupling opens the way to control energy which can be transferred to (heating) or removed from (cooling) the phonon reservoir depending on the parameters of the electromagnetic pulse.

The evident example of a two-level dissipative system (TLDS) is a semiconductor quantum dot (QD) which is widely used in various applications of the quantum optics, nanoelectronics, and quantum information. Physical features of QD based setups are ultimately depend on the interaction between the electron system and the phonon reservoir. This interaction is responsible for changing the quantum states of the phonon subsystem and, thus, the temperature of the QD. From a practical point of view, it is important to control the energy transfer between the electron and phonon subsystems of the QD. For instance, the processes of the coherent phonon generation [1] and cooling [2] can be implemented using the optical control. One of the convenient approaches to organize such control is the dynamical Stark shift method. This method has been used to realize the fast quantum gates for trapped cold ions [3], the atomic cooling [4], the precise qubit thermometry of an oscillator [5], the population inversion of a single QD [6], and the coherent quantum interference between the Stark split dressed states (DSs) in a semiconductor quantum well [7].

Interaction of a dissipative system with a strong electromagnetic field leads to changing both the energy spectrum of electron states [8, 9] and the energy spectrum of vibrational modes of the reservoir [10]. Thus, an additional mode appears in the vibrational spectrum due to the periodic displacement of equilibrium positions of reservoir oscillators under the action of the Rabi oscillations of the electron density [11, 12]. Additionally, the optically coupled electron levels become split in the conditions of the ac Stark effect. This fact is known as removing the degeneracy in the electron-photon system. Since the energy separation between the Stark split levels is comparable with the energy of vibrational modes, the nonadiabatic effects must be taken into account [13]. It is worth to stress that the considered physical situation can be interpreted as a manifestation of the pseudo-Jahn-Teller effect [14] in the electron-photon system. In this Letter, the effect of the displacement of the equilibrium position (EDEP) of vibrational coordinates under the action of a strong electromagnetic field is taking into account for describing the energy transfer between the two-level system and the phonon reservoir.

Consider a two-level electron subsystem interacting with a strong monochromatic electromagnetic field and a phonon reservoir. Let us divide the total Hamiltonian into two parts. First part, $H_1$, contains the electron and photon subsystems and their interaction in the dipole approximation and the rotating wave approximation (RWA) while second part, $H_2$, includes the energy of the phonon subsystem, $H_{pn}^0$, and the energy of the electron-phonon interaction, $V_{e\text{-pn}}$. Thus, the Hamiltonian of the system can be written as

$$H = H_1 + H_2, \quad H_1 = \frac{\varepsilon}{2}\sigma_z + \hbar\omega_0 c^+ c + \hbar g_0 \sigma_+ c + \hbar g_0^* c^+ \sigma_-,$$
$$H_2 = H_{pn}^0 + V_{e\text{-pn}}. \quad (1)$$

Here, $\varepsilon$ is the energy difference between the ground and excited electron levels; $\sigma_z$, $\sigma_+$, and $\sigma_-$ are the Pauli operators associated to the TLDS with the ground $|\varphi_1\rangle$ and the excited $|\varphi_2\rangle$ electron states; $c$ and $c^+$ are the annihilation and creation operators of the electromagnetic field of frequency $\omega_0$; $g_0$ is the coupling strength of the electron-photon interaction.

The solution of the Schrödinger equation for $H_1$ is well known and can be written in the basis of electron states that are dressed by an electromagnetic field. If initially only the ground electron state is populated, the solution has the following form [15]

$$|\Psi(t),m,n\rangle = C_m U_m^+ |\xi_1(t),m,n\rangle + C_m S_m U_m^- |\xi_2(t),m,n\rangle,$$

$$U_m^\pm = \sqrt{\frac{1}{2}\left(1\pm\frac{\delta}{2\Omega}\right)}, \; S_m = \frac{g_0}{|g_0|}, \; 2\Omega = \sqrt{\delta^2 + 4|g_0|^2(m+1)},$$

$$\delta = \hbar^{-1}\varepsilon - \omega_0. \quad (2)$$

Here, $m$ and $n$ are the numbers of photons and phonons, respectively; $C_m$ is the initial probability amplitude for the state with $m$ photons; $\Omega$ is the off-resonance Rabi frequency; $\delta$ is the detuning of the optical field with respect to the electron transition frequency. It should be stressed that the wave function in Eq. (2) is defined for the instant start of the electron-photon interaction in the so-called diabatic case [15]. Including $H_1$ into the zeroth-approximation Hamiltonian of $H_2$, the matrix element of the electron-vibrational interaction can be defined as

$$H' = \langle \Psi(t),m,n|\tilde{V}_{\text{e-pn}}|\Psi(t),m,n\rangle,$$
$$\tilde{V}_{\text{e-pn}} = \exp(-i\hbar^{-1}H_0 t) V_{\text{e-pn}} \exp(i\hbar^{-1}H_0 t),$$
$$H_0 = H_1 + H_{\text{pn}}^0. \quad (3)$$

The energy difference between the DSs is $2\hbar\Omega$. For the dynamical Stark shift experiments, this difference is comparable with the vibrational energy of reservoir modes. Such situation gives rise to existence of the DSs which are mixed by the electron-vibrational interaction. In order to take into account the EDEP of vibrational coordinates, let us consider the non-diagonal matrix elements in the DS basis and transfer from this basis to the basis of the bare states (BSs), using the following relationships

$$|\xi_1(t),m,n\rangle = U_m^+ e^{i\left(\Omega-\frac{\delta}{2}\right)t} |\varphi_1(t),m+1,n\rangle$$
$$-S_m U_m^- e^{i\left(\Omega+\frac{\delta}{2}\right)t} |\varphi_2(t),m,n\rangle,$$

$$|\xi_2(t),m,n\rangle = S_m^* U_m^- e^{-i\left(\Omega+\frac{\delta}{2}\right)t} |\varphi_1(t),m+1,n\rangle$$
$$+U_m^+ e^{-i\left(\Omega-\frac{\delta}{2}\right)t} |\varphi_2(t),m,n\rangle. \quad (4)$$

Thus, the non-diagonal part of the $H'$ in the DS basis can be written in the form (for details see Ref. [16])

$$H'_{\text{nd}} = H'_{12} + H'_{21} = 2\left(C_m U_m^+ U_m^-\right)^2 \left(\tilde{V}_{11} - \tilde{V}_{22} - \Delta\tilde{V}_{22}\right)\cos(2\Omega t)$$
$$+ C_m^2 S_m U_m^+ U_m^- \left[(U_m^+)^2 e^{-i2\Omega t} - (U_m^-)^2 e^{i2\Omega t}\right]\left(\tilde{V}_{12} - \Delta\tilde{V}_{12}\right)e^{i\delta t}$$
$$- C_m^2 S_m^* U_m^+ U_m^- \left[(U_m^+)^2 e^{i2\Omega t} - (U_m^-)^2 e^{-i2\Omega t}\right]\left(\tilde{V}_{21} - \Delta\tilde{V}_{21}\right)e^{-i\delta t},$$

$$\tilde{V}_{ij} = \langle \varphi_i(t),m,n|\tilde{V}_{\text{e-pn}}|\varphi_j(t),m,n\rangle. \quad (5)$$

Here, $\Delta\tilde{V}_{ij}$ are the terms that describe the contribution of the EDEP to the electron-vibrational interaction. For further consideration, we leave only these terms and determine them up to the first order with respect to the small displacement between two equilibrium positions of oscillations of the normal coordinates related to the ground and excited BSs. In this approximation, the electron-vibrational interaction can be represented in the following form [16]

$$H'_{\text{ED}} = \hbar F_0 e^{-i2\Omega t} + \hbar F_0^* e^{i2\Omega t},$$

$$F_0 = \sum_{\mathbf{k},s} \hbar f_{\mathbf{k},s} U_m^+ U_m^- C_m^2 \left[(U_m^+)^2 S_m g_{12}^{\mathbf{k},s} - (U_m^-)^2 S_m^* g_{21}^{\mathbf{k},s}\right.$$
$$\left. - 2U_m^+ U_m^- g_{22}^{\mathbf{k},s}\right], \; f_{\mathbf{k},s} = \sum_{l,\alpha} \frac{\Delta_{l,\alpha} \cdot \mathbf{e}_{\mathbf{k},s}^\alpha}{\sqrt{2N m_\alpha \hbar \omega_{\mathbf{k},s}}} e^{i\mathbf{k}\cdot\mathbf{R}_{l,\alpha}^g},$$

$$\Delta_{l,\alpha} = \mathbf{R}_{l,\alpha}^e - \mathbf{R}_{l,\alpha}^g. \quad (6)$$

Here, $s$ denotes the phonon mode of frequency $\omega_{\mathbf{k},s}$ and wave vector $\mathbf{k}$; $g_{ij}^{\mathbf{k},s}$ is the coupling strength of the electron-phonon interaction; $\mathbf{e}_{\mathbf{k},s}^\alpha$ are the normalized polarization vectors of vibrational coordinates for an $\alpha$-type oscillator of mass $m_\alpha$; $N$ is the number of reservoir oscillators; $\Delta$ is the displacement operator; $\mathbf{R}^g$ and $\mathbf{R}^e$ are the equilibrium positions of oscillations of the normal coordinates related to the ground and excited BSs, respectively. The obtained matrix element is responsible for establishment of the photoinduced vibronic (PIV) coupling in the TLDS. The matrix element describes the vibrations of frequency $2\Omega$ under the action of the Rabi oscillations of the electron density between the BSs. These vibrations have the amplitude which is proportional to the module of the displacement, $\Delta$, and the coupling constants of the electron-vibrational and electron-photon (factor $U_m^+ U_m^-$) interactions. Since the frequency of the vibrations is equal to the frequency of the electron transition between the DSs, the electron-vibrational interaction leads to the electron density oscillations between these states and defines the back action of the vibrations on the electron density. Thus, the interaction of Eq. (6) gives rise to appearance of the vibronic coupling [17] in the electron-photon system.

To determine the PIV states which can be represented as a linear combination of the DSs, we have to solve the following differential system for coefficients

$$\frac{\partial}{\partial t} d_{1,n}(t) = -iF_0 e^{-i2\Omega t} d_{2,n}(t),$$
$$\frac{\partial}{\partial t} d_{2,n}(t) = -iF_0^* e^{i2\Omega t} d_{1,n}(t). \quad (7)$$

The solution of the system gives the PIV wave functions for different initial conditions. Moreover, using this solution, we can write the PIV states in three different bases (i) in the BS basis, (ii) in the basis of the states which are dressed by the electron-photon interaction, and (iii) in the basis of the doubly dressed states (DDSs), i.e. the states which are dressed firstly by the electron-photon interaction and, then, by the electron-vibrational interaction (Fig. 1). In the last case, the PIV wave function takes the following form

$$|\nu(t),m,n\rangle = \frac{1}{\sqrt{2}}|\zeta_1(t),m,n\rangle + \frac{1}{\sqrt{2}}|\zeta_2(t),m,n\rangle,$$

$$|\zeta_1(t),m,n\rangle = C_m U_m^+ W_1^+ e^{-i(2\Omega+\Omega_1)t}|\xi_1(t),m,n\rangle$$
$$+ C_m S_m U_m^- W_2^+ e^{i(2\Omega-\Omega_1)t}|\xi_2(t),m,n\rangle,$$
$$|\zeta_2(t),m,n\rangle = C_m U_m^+ W_1^- e^{-i(2\Omega-\Omega_1)t}|\xi_1(t),m,n\rangle$$
$$+ C_m S_m U_m^- W_2^- e^{i(2\Omega+\Omega_1)t}|\xi_2(t),m,n\rangle,$$
$$W_1^\pm = \frac{1}{\sqrt{2}} \pm \frac{U_m^- F_0 - U_m^+ \Omega}{\sqrt{2} U_m^+ \Omega_1},\quad W_2^\pm = \frac{1}{\sqrt{2}} \pm \frac{U_m^+ F_0^* + U_m^- \Omega}{\sqrt{2} U_m^- \Omega_1},$$
$$\Omega_1 = \sqrt{\Omega^2 + |F_0|^2}.\quad (8)$$

The initial conditions for the coefficients $d_{1,n}$ and $d_{2,n}$ correspond with the electron population of the DSs in the electron-photon system [see Eq. (2)]. Eqs. (8) show that the electron population in the TLDS oscillates between the DSs with the frequency $\Omega_1$ and becomes equally divided between the DDSs. At the same time, the phonon numbers are not changed under the action of the considered interaction contribution.

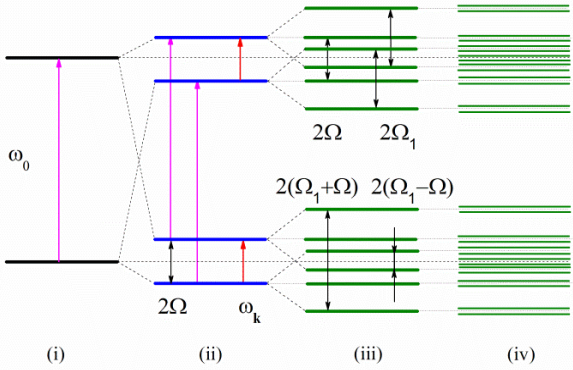

FIG. 1. (Color online) Formation of the DDSs in the TLDS: (i) BSs with an electromagnetic field of frequency $\omega_0$ (magenta, dashed arrows); (ii) DSs with a phonon mode of frequency $\omega_k$ (red, doted arrows); (iii) DDSs in the first order of the electron-vibrational interaction; (iv) DDSs in the second order of the electron-vibrational interaction.

In order to establish the coupling with the phonon modes, we must take into account the next term of the Taylor's series expansion in displacements for the interaction operator. Using the wave functions of Eqs. (8), the matrix element of the quadratic interaction can be readily obtained (for details see Ref. [16])

$$H''_{ED} = \frac{\hbar C_m^2}{2}\Big[\big(W_1^{+*}W_2^+ + W_1^{-*}W_2^-\big)e^{i2\Omega t} + W_1^{+*}W_2^- e^{i2(\Omega+\Omega_1)t}$$
$$+ W_1^{-*}W_2^+ e^{i2(\Omega-\Omega_1)t}\Big]$$
$$\times \sum_{k,s,n}\Big\{\big(F_{11}^{k,s}+F_2^{k,s}\big)\sqrt{n_{k,s}}\,e^{-i\omega_{k,s}t} + \big(F_{12}^{k,s}-F_2^{k,s}\big)\sqrt{n_{k,s}+1}\,e^{i\omega_{k,s}t}\Big\}$$

$$+ \frac{\hbar C_m^2}{2}\Big[\big(W_2^{+*}W_1^+ + W_2^{-*}W_1^-\big)e^{-i2\Omega t} + W_2^{+*}W_1^- e^{-i2(\Omega+\Omega_1)t}$$
$$+ W_2^{-*}W_1^+ e^{-i2(\Omega-\Omega_1)t}\Big]$$
$$\times \sum_{k,s,n}\Big\{\big(F_{12}^{k,s}-F_2^{k,s*}\big)\sqrt{n_{k,s}}\,e^{-i\omega_{k,s}t} + \big(F_{11}^{k,s}+F_2^{k,s*}\big)\sqrt{n_{k,s}+1}\,e^{i\omega_{k,s}t}\Big\},$$

$$F_2^{k,s} = \hbar^2 f_{k,s} U_m^+ U_m^-\left(n_{k,s}+\frac{1}{2}\right)\Big[(U_m^+)^2 S_m h_{12}^{k,s} + (U_m^-)^2 S_m^* h_{21}^{k,s}\Big],$$
$$F_{11}^{k,s} = \hbar^2 f_{k,s} U_m^-(U_m^+)^2\Big(U_m^- h_{22}^{k,s} - U_m^+ S_m h_{12}^{k,s}\Big),$$
$$F_{12}^{k,s} = \hbar^2 f_{k,s} U_m^+(U_m^-)^2\Big(U_m^+ h_{22}^{k,s} + U_m^- S_m h_{12}^{k,s}\Big).\quad (9)$$

Here, $n_{k,s}$ is the phonon number of mode $s$ and wave vector $k$; $h_{ij}^{k,s}$ is the quadratic coupling strength of the electron-phonon interaction. In Eq. (9), we have left only the non-diagonal matrix elements in the basis of the DSs. It can be seen from Eq. (9) that there are three frequencies in the PIV state spectrum, $2\Omega$, $2(\Omega-\Omega_1)$, and $2(\Omega+\Omega_1)$. Note that the diagonal matrix elements in the basis of the DSs give additional frequency, $2\Omega_1$ (see also Fig. 1). Such situation is similar to existence of a Mollow triplet in the electron-photon system. In further consideration, we will take into account only the term which defines the transitions of frequency $2\Omega$. Other terms can be accounted in the same manner. Then, we assume that the parameters of the electromagnetic field are chosen so that there is a resonance between the transition frequency $2\Omega$ of the DSs and one reservoir phonon mode ($s$, $k_0$). This assumption makes it possible to use the RWA for the phonon mode.

For each selected mode ($s$, $k_0$), the Hamiltonian associated with Eq. (9) is the Jaynes-Cummings interaction Hamiltonian. Since this Hamiltonian commute with the parity operator [18], $\Pi = -\sigma_z \exp(i\pi\hat{n})$ where $\hat{n}$ is the phonon number operator, the final form of the PIV wave functions in the DSs basis can be written as [19]

$$|\nu_-(t),m,n\rangle = D_{1,2n}^-(t)|\xi_1,m,2n\rangle + D_{2,2n+1}^-(t)|\xi_2,m,2n+1\rangle,$$
$$|\nu_+(t),m,n\rangle = D_{1,2n-1}^+(t)|\xi_1,m,2n-1\rangle$$
$$+ D_{2,2n}^+(t)|\xi_2,m,2n\rangle.\quad (10)$$

Here, signs + and − stand for the even and odd parity, respectively. Strictly speaking, the BS wave functions also have the parity according to the number of photons but we omit this fact for simplicity. Using the RWA, the coefficients of Eqs. (10) can be found from the following differential systems

$$\frac{\partial}{\partial t}D_{1,2n-0.5\mp0.5}^\pm(t) = -i G_{k_0}^\pm e^{i\delta_1 t} D_{2,2n+0.5\mp0.5}^\pm(t),$$
$$\frac{\partial}{\partial t}D_{2,2n+0.5\mp0.5}^\pm(t) = -i G_{k_0}^{\pm*} e^{-i\delta_1 t} D_{1,2n-0.5\mp0.5}^\pm(t),$$
$$G_{k_0}^\pm = \frac{C_m^2}{2}\sqrt{2n_{k_0}+\frac{1}{2}\mp\frac{1}{2}}\big(F_{11}^{k_0}+F_2^{k_0}\big)\big(W_1^{+*}W_2^+ + W_1^{-*}W_2^-\big),$$
$$\delta_1 = 2\Omega - \omega_{k_0}.\quad (11)$$

The solution of Eqs. (11) allows to write the final PIV wave functions of the corresponding parity in three different bases in the same manner as the wave functions of Eqs. (8). In the case of the DDS basis, the energy eigenvalues take the form similar as in the DS case

$$E_1^\pm = \hbar\omega_0(m+1) + \hbar\omega_{\mathbf{k}_0}\left(2n_{\mathbf{k}_0} \mp \frac{1}{2}\right) - \hbar\Omega_2^\pm,$$

$$E_2^\pm = \hbar\omega_0(m+1) + \hbar\omega_{\mathbf{k}_0}\left(2n_{\mathbf{k}_0} \mp \frac{1}{2}\right) + \hbar\Omega_2^\pm,$$

$$\Omega_2^\pm = \sqrt{\delta_1^2 + 4\left|G_{\mathbf{k}_0}^\pm\right|^2}. \quad (12)$$

For an exact resonance, the difference of module squares of the coefficients can be written as

$$\left|D_{1,2n-0.5\mp 0.5}^\pm(t)\right|^2 - \left|D_{2,2n+0.5\mp 0.5}^\pm(t)\right|^2 = \left[(\bar{U}_m^+)^2 - (\bar{U}_m^-)^2\right]$$
$$\times \cos(2\Omega_{20}^\pm t) - i\bar{U}_m^+\bar{U}_m^-\left(S_m^* s_n^* - S_m s_n\right)\sin(2\Omega_{20}^\pm t),$$

$$s_n = \frac{G_{\mathbf{k}_0}}{|G_{\mathbf{k}_0}|},\ \bar{U}_m^+ = U_m^+ C_m D_{1,2n-0.5\mp 0.5}^\pm(0),$$

$$\bar{U}_m^- = S_m U_m^- C_m D_{2,2n+0.5\mp 0.5}^\pm(0). \quad (13)$$

Here, $\Omega_{20}$ is the resonance frequency of $\Omega_2$ with $\delta_1 = 0$. In Eq. (13), we assume that the TLDS state is initially a product state which is composed by its subsystem states. Namely, in the equations above, $C_m$ and $D_{i,n}^\pm(0)$ are the initial probability amplitudes for the photon and phonon states, respectively. Eq. (13) shows that the initially presented population difference between the DSs in the electron-photon system oscillates with the frequency of $2\Omega_{20}$.

Thus, the DDSs are formed by two steps. At the first step, the basis of levels is created due to the first order electron-vibrational interaction. At the second step, the second order interaction leads to broadening of the levels. Finally, taking into account the dispersion of phonon states, we obtain a quasi-continuous electron spectrum in the vicinity of the DSs (Fig. 1).

In order to include the dissipation in our consideration, we have to transfer from the wave function description to the density matrix formalism and specify the relaxation channels. The population of the DSs is affected by both the nonradiative and radiative decays. In both cases, we assume that the Weisskopf-Wigner approximation is valid and the phenomenological damping rates can be introduced. This assumption leads to the complete equivalence of the DS and DDS representations (see Table I). The exact estimations of the nonradiative decay rate must be done using the coupling between the photoinduced vibrations and phonon modes of the reservoir [see Eq. (9)] but it is not the aim of this Letter. Additionally, in the case of the ac Stark effect, we assume that the Rabi frequency is large compared to the radiative decay rate.

TABLE I. Equivalence of representations of the dressed and doubly dressed states.

| Electron-photon interaction | Electron-phonon interaction |
| --- | --- |
| Electromagnetic field mode of frequency $\omega_0$ | Phonon mode of frequency $\omega_{\mathbf{k}}$ |
| Oscillations between the BSs of frequency $\Omega$ | Oscillations between the DSs of frequency $\Omega_1$ |
| Appearance of the DSs | Appearance of the DDSs |
| Radiative decay due to the interaction with the vacuum field mode continuum | Nonradiative decay due to the interaction with the phonon mode continuum |

For our consideration, existence of the nonradiative decay means the uniform distribution of the electron population, oscillating between the DSs. Thus, if initially the lower DS has the larger population than the upper DS, the energy will be removed from the phonon reservoir, in the opposite case, if initially the lower DS has the lower population than the upper DS, the energy will be transferred to the phonon reservoir. In the case of the equal populations at the DSs, the energy is not changed. Transferred energy can be defined as the product of the electron population difference and the energy difference of the DSs and equals

$$2\hbar\Omega\left((U_m^+)^2 - (U_m^-)^2\right) = \hbar\delta, \quad (14)$$

in accordance with the energy conservation law.

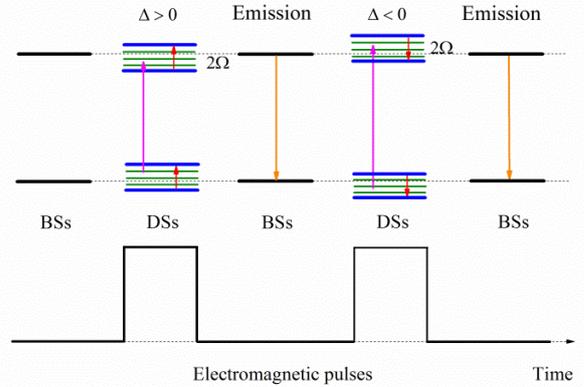

FIG. 2. (Color online) Energy transfer using the dynamical Stark shift in the TLDS: reservoir cooling ($\Delta > 0$) and reservoir heating ($\Delta < 0$).

It is important to stress that the considered transference of energy can be observed if the frequency of the photoinduced oscillations are comparable with the frequencies of vibrational modes of the reservoir. To satisfy this condition, we must use the optical detunings of the same order of magnitude as the resonance Rabi frequency. Thus, we can control the energy transfer between the electron and phonon subsystems with use of the dynamical Stark shift, choosing properly the intensity and optical frequency of an electromagnetic pulse. Appling a sequence of pulses with the duration that is shorter compared to the radiative decay time, we can heat (negative optical detuning) or cool (positive optical detuning) the TLDS (Fig. 2).

In conclusion, we describe the mechanism of energy transfer between the two-level electron system and the phonon reservoir in the case of the RWA. The use of the RWA means the weak electron-vibrational coupling in the TLDS. Remarkably, there are no limitations for this mechanism to occur in the case of the strong coupling. In the strong coupling case, the generalized RWA may be used [18, 20]. At the same time, the case of low temperatures, when some phonon modes are frozen, has to be considered separately. It should be noted that the vibronic coupling may be created by the interaction between optical phonon modes and electron states directly [16] as it occurs in the doped solid state systems [19, 21]. Therefore, the results of this work are also valid in the case of optical phonon modes. Since the exact resonance conditions must be satisfied for the optical phonon frequency and the transition frequency between the DSs, the case of the forced vibrations of reservoir oscillators considered in this Letter is the most general.

The author of this Letter has been inspired by the recently published experimental results of Ref. [22].

Financial support of the Ministry of Education and Science of the Russian Federation (grant 074-U01) and of the Russian Foundation for Basic Research (grant 17-02-00598) are gratefully acknowledged.

*e-mail address: avivanov@mail.ifmo.ru